\documentclass[aps,prd,showkeys,showpacs,amssymb,twocolumn,
amsfonts,epsf,preprintnumbers,nofootinbib,superscriptaddress]{revtex4}

\usepackage[dvips]{graphicx}
\usepackage{bm,latexsym,amsmath,amssymb,amsfonts,color}

\def\be {\begin{equation}}
\def\ee {\end{equation}}
\def\ba {\begin{eqnarray}}
\def\ea {\end{eqnarray}}

\def\bi {\begin{itemize}}
\def\ei {\end{itemize}}
\newcommand\beq{\begin{eqnarray}}
\newcommand\eeq{\end{eqnarray}}
\newcommand{\bea}{\begin{eqnarray}}
\newcommand{\eea}{\end{eqnarray}}

\usepackage{graphicx}
\usepackage{dcolumn}
\usepackage{bm,morefloats,color}

\usepackage{amssymb,bm}

\def\X5sp{{\rm X}_5}
\def\Y3sp{{\rm Y}_3}
\def\Z3sp{{\rm Z}_3}

\begin{document}

\title{Disformal transformation of cosmological perturbations}

\author{Masato Minamitsuji}
\affiliation{Multidisciplinary Center for Astrophysics (CENTRA), Instituto Superior T\'ecnico, Lisbon 1049-001, Portugal.}

\begin{abstract}
We investigate the gauge-invariant cosmological perturbations in the gravity and matter frames in the general scalar-tensor theory where two frames are related by the disformal transformation. The gravity and matter frames are the extensions of the Einstein and Jordan frames in the scalar-tensor theory where two frames are related by the conformal transformation, respectively. First, it is shown that the curvature perturbation in the comoving gauge to the scalar field is disformally invariant as well as conformally invariant, which gives the predictions from the cosmological model where the scalar field is responsible both for inflation and cosmological perturbations. Second, in case that the disformally coupled matter sector also contributes to curvature perturbations, we derive the evolution equations of the curvature perturbation in the uniform matter energy density gauge from the energy (non)conservation in the matter sector, which are independent of the choice of the gravity sector. While in the matter frame the curvature perturbation in the uniform matter energy density gauge is conserved on superhorizon scales for the vanishing nonadiabatic pressure, in the gravity frame it is not conserved even if the nonadiabatic pressure vanishes. The formula relating two frames gives the amplitude of the curvature perturbation in the matter frame, once it is evaluated in the gravity frame.
\end{abstract} 
\pacs{04.50.Kd, 98.80.-k}
\keywords{Modified theories of gravity, Cosmology}

\date{\today}

\maketitle

\section{Introduction}

The Concordance Model of Cosmology 
has succeeded in explaining the history of the universe \cite{obs}.
However, we still do not understand which fundamental physics is 
behind the elements of the Concordance Model, i.e., inflation, dark matter and dark energy.
In the last decades, cosmologists and gravitational physicists
have explored the possible modifications of the Einstein gravity on cosmological scales \cite{mod1},
as the alternatives to these elements.
A naive modification of the Einstein gravity provides ghostlike degrees of freedom arising from the higher derivative terms
as well as inconsistencies with experimental tests on the Einstein gravity.
To avoid the appearance of the ghostlike degrees of freedom associated with Ostrogradski's theorem \cite{ostro}, 
the equations of motion should be written in terms of the second order differential equations.
On the other hand,
to pass the experimental tests on the Einstein gravity,
a realistic modification of gravity should contain a mechanism to suppress scalar interactions on small scales \cite{sc1}.
After a number of models have been examined,
it has turned out that successful models of the modified gravity
can be described in terms of a class of Horndeski's scalar-tensor theory.
The theory was originally proposed by Hordenski \cite{hor1} forty years ago,
and recently has been reformulated with the growing interest in applications to  cosmological problems \cite{hor2}.
Horndeski's theory is known as the most general scalar-tensor theory
where the equations of motion remain of the second order,
despite the existence of the derivative interactions.
Horndeski's theory has been investigated from the various cosmological aspects,
e.g., dark energy \cite{hor_de1}, inflation \cite{hor_inf1}, early universe \cite{hor_inf2},
screening mechanisms \cite{screening1} and also observational constraints \cite{constr1}.

As the generalization, 
we may consider the situation that 
the scalar field is directly coupled to the matter sector. 
In such a theory,
matter does not follow geodesics associated with the gravitational metric $g_{\mu\nu}$
but that associated with the other metric ${\bar g}_{\mu\nu}$
which differs from $g_{\mu\nu}$ by the contributions of the scalar field.
The most familiar and well-studied case with two different metrics for
gravity and matter is that
the matter frame metric ${\bar g}_{\mu\nu}$ can be constructed by the gravitational one $g_{\mu\nu}$
and the scalar field $\phi$ itself, but not by the derivatives of the scalar field.
In this case,
the matter frame metric is  
conformally related to the gravity frame one by $\bar g_{\mu\nu}=\alpha (\phi)g_{\mu\nu}$ \cite{mod1}.
The gravity and matter frames, $g_{\mu\nu}$ and ${\bar g}_{\mu\nu}$ ,
 are often referred to as the Einstein and Jordan frames, respectively.
The conformal transformation does not modify the causal structure of the spacetime.
Concerning cosmological perturbations,
after a number of studies \cite{cf1,cf3,cf4,cf5},
it has been shown that 
in the case that a single scalar field is responsible for inflation and cosmological perturbations
the curvature perturbation in the comoving / uniform energy density gauge,
which is conserved on superhorizon scales,
is conformally invariant for all orders of perturbations,
allowing us to evaluate it in the most convenient Einstein frame,  
even though the physical frame for matter is the Jordan frame.
By the ``physical frame'' , we mean the frame in which matter minimally couples to
its metric and  photons propagate on null geodesics associated with it.

As the most general case of the scalar-tensor theory with two different metrics for gravity and matter, however,
we may consider the matter frame metric 
which can also be constructed from the derivatives of the scalar field
as well as the gravitational metric and the scalar field itself, 
${\bar g}_{\mu\nu}={\bar g}_{\mu\nu} (g_{\mu\nu},\phi,\partial\phi,\partial^2\phi,\cdots)$.
As the next simplest case to the conformal transformation,
it would be reasonable to truncate the expansion of ${\bar g}_{\mu\nu}$ with respect to the order of the scalar field derivatives 
at the first order, 
by assuming that the effects of the higher derivatives would be suppressed by the higher inverse powers of the cut-off mass scale
below which an effective theory description is assumed to be valid. 
Although the expansion argument does not distinguish 
operators of $(\partial \phi)^n$ from those of $\partial^n\phi$ ($n\geq 2$),
we would keep only the former one,
because the latter would give the higher derivative terms in the equations of motion 
and hence give rise to ghostlike instabilities associated with Ostrogradski's theorem.  
Then according to \cite{diff1},
the next simplest matter frame metric involves the first order derivatives of the scalar field as well as the scalar field itself:
\bea
\label{transfor0}
{\bar g}_{\mu\nu}=\alpha (X, \phi)g_{\mu\nu}+ \beta(X, \phi)\phi_{\mu}\phi_{\nu},
\eea
where $\phi_\mu=\nabla_\mu\phi$ is the covariant derivative of the scalar field associated the gravity frame metric $g_{\mu\nu}$
and $X:=-\frac{1}{2}g^{\mu\nu}\phi_\mu\phi_\nu$.
Eq. \eqref{transfor0} is often called the disformal transformation.
In \cite{diff1},
in the context of the Finsler geometry
how in general the gravity and matter frames are related by a single scalar field
was argued, and it was shown that its reduction to the Riemannian geometry has to be given by the disformal relation \eqref{transfor0}. 
As pointed out in \cite{rule1}, however,
the disformal transformation \eqref{transfor0} of a class of Horndeski's theory
could give the higher derivative coupling terms which are absent in Horndeski's theory,
and hence the higher derivative terms in the equations of motion. 
Nevertheless, as argued  in \cite{rule2},
the appearance of Ostrogradski's ghosts may be able to be avoided by the implicit constraints,
implying the existence of the more general healthy scalar-tensor theory beyond Horndeski
\footnote{See also \cite{rule3} for an explicit construction of the scalar-tensor theory
beyond Horndeski.}.
In this paper, however,  we will focus on the simpler case that the disformally-transformed gravitational theory 
also belongs to a class of Horndeski's theory \cite{rule1}, 
namely
\bea
\label{transfor}
{\bar g}_{\mu\nu}=\alpha (\phi)g_{\mu\nu}+ \beta(\phi)\phi_{\mu}\phi_{\nu}.
\eea
Such a simplification is made,
in order to restrict our arguments to Horndeski's scalar-tensor theory.
For $\beta=0$, \eqref{transfor} reduces to the ordinary conformal transformation,
while when $\alpha=1$, $\beta$ represents the pure disformal transformation.
First, we impose that $\alpha>0$ so that for $\beta=0$ the conformal transformation is well-defined.
In contrast to the conformal transformation, the disformal transformation may change the causal structure of spacetime.
Consider a null vector field $v^{\mu}$ for the matter frame metric $\bar g_{\mu\nu}$,
$\bar g_{\mu\nu}v^\mu v^\nu=0$,
which can be rewritten as $g_{\mu\nu}v^\mu v^\nu =-\frac{\beta}{\alpha} (\phi_\mu v^\mu)^2$.
Thus
for $\beta>0$, $g_{\mu\nu}v^\mu v^{\nu}<0$ and hence $v^{\mu}$ becomes a timelike vector field for the gravity frame metric $g_{\mu\nu}$,
and 
for $\beta<0$, $g_{\mu\nu}v^\mu v^{\nu}>0$ and hence $v^{\mu}$ becomes a spacelike vector field for $g_{\mu\nu}$.
The matter frame metric with the upper indices is given by
\bea
\label{inverse}
{\bar g}^{\mu\nu}
=\frac{1}{\alpha}
\Big(g^{\mu\nu}-\frac{\beta}{\alpha-2\beta X}\phi^{\mu}\phi^{\nu}\Big).
\eea
In order to obtain a healthy disformal transformation
we should impose the following conditions:

\vspace{0.1cm}

(1)  In order for ${\bar g}_{\mu\nu}$ to have the Lorentzian signature ${\bar g}_{00}<0$, $\alpha g_{00}+\beta \phi_0^2<0$.

\vspace{0.1cm}

(2) In order for ${\bar g}_{\mu\nu}$ to be causal $d\bar s^2= {\bar g}_{\mu\nu}dx^\mu dx^\nu<0$, $\beta<0$. 

\vspace{0.1cm}

(3) In order for \eqref{inverse} also to be well-defined and 
 have the Lorentzian signature with ${\bar g}^{00}<0$, $\alpha-2\beta X>0$.

\vspace{0.1cm}

These conditions have been argued in \cite{rule1} (see also \cite{diff1}).
Applications of the disformal transformation to the cosmological problems
have been argued, 
such as to inflation \cite{disf1},
cosmology with varying speed of light \cite{vsl},
dark energy \cite{de},
screening mechanism \cite{dis_scr0,dis_scr,dis_scr2},
MOND \cite{mond},
dark matter \cite{dm}
and observational constraints \cite{obs2}.

The purpose of this paper is to clarify how the gauge-invariant cosmological  perturbation variables
\cite{mfb,ks} in the gravity and matter frames
are related by the disformal transformation \eqref{transfor}.
To our knowledge, so far there has been no formulation 
of relating the gauge-invariant perturbation variables in both frames. 
First, we investigate whether the curvature perturbation in the comoving gauge to the scalar field 
is invariant under the disformal transformation
as well as the conformal transformation  \cite{cf1,cf3,cf4}.
In the simplest case where there is no matter field disformally coupled to gravity
and the only scalar field is responsible both for inflation and cosmological perturbations,
the comoving curvature perturbation is conserved after the scale of interest crosses the horizon,
which gives the final prediction from the given model.
Thus if it is disformally invariant, 
it may be evaluated in any disformally related frame 
as done in the Einstein frame in the gravitational theory with the nonminimal coupling of
the scalar field to the Ricci scalar
where two frames are related by the conformal transformation.

Second, we will consider the case where there is 
the matter sector disformally coupled to the scalar field.
In such a case,
the curvature perturbation will not be conserved due to the continual conversion of the isocurvature perturbation between the scalar field and matter
to the curvature perturbation
and even the fluctuations of the disformally coupled matter
may be the dominant sources of cosmological perturbations, instead of the scalar field itself.
To obtain predictions from such a model,
we need to solve the evolution equation of the curvature perturbation on the uniform matter energy density gauge.
A model to be of use as a possible reference was considered in Ref. \cite{disf1}.
In this model, the gravity frame experiences a decelerating expansion 
driven by a canonical kinetic term of the scalar field
while the disformally related matter frame experiences a short inflationary expansion in the early phase,
and the fluctuations of the scalar field in the decelerating gravity frame
could not produce the realistic scale-invariant spectrum of the curvature perturbation.
Instead, the disformally coupled matter quantized in the inflationary matter frame 
may be able to source the curvature perturbation via a curvaton-like mechanism \cite{curv1}.
Although in this paper we will not consider a particular model,
such a model may be able to be generalized to the case of a class of Horndeski's theory.
As the first step to investigate the evolution of the curvature perturbations in such a model,
it will be important to formulate how the curvature perturbation in the uniform matter energy density gauge evolves
in the presence of the disformal coupling.
Following \cite{malik}, the evolution equation of the curvature perturbation in the uniform matter energy density gauge
will be derived as the consequence of the energy (non)conservation in the matter sector,
which will be independent of the choice of the gravity sector.
Therefore, we will derive the gauge-invariant evolution equations of the curvature perturbation in the uniform energy density
of each component in the gravity and matter frames.

The construction of the paper is as follows.
In Section II, we review the covariant equations of motion in the gravity and matter frames in the general scalar-tensor theory
with the matter sector disformally coupled to the scalar field.
In Section III, we define the gauge-invariant gravitational perturbation variables in the gravity and matter frames
and derive the relations between the corresponding perturbation variables in both frames.
In Section IV, similarly, we define the gauge-invariant matter perturbation variables in the gravity and matter frames
and derive the relations between the corresponding perturbation variables in both frames.
In Section V, 
we derive the evolution equations of the curvature perturbations in the uniform matter energy density gauges
in both frames.
The paper is closed after giving a brief summary in Section VI.

\section{The covariant equations of motion in the gravity and matter frames}

We consider the scalar-tensor theory with matter disformally coupled to the scalar field
\bea
\label{lag1}
S&=&S_g+ S_m; 
\nonumber\\
S_g&:=& \int d^4 x \sqrt{-g}{\cal L}_g[g,\phi] ,
\nonumber\\
S_m&:=&\int d^4 x \sqrt{-\bar g}{\cal L}_m [\bar g,\Psi],
\eea
where $g_{\mu\nu}$ is the gravity frame metric,
${\cal L}_g$ is the Lagrangian of the gravity sector given by Horndeski's theory,
composed of the four independent combinations of the scalar field operators:
\bea
{\cal L}_g[g,\phi]= \sum_{i=2}^5 {\cal L}_i[g,\phi],
\eea 
with
\bea
\label{gal_acc}
{\cal L}_2&=& P(X,\phi),
\quad
{\cal L}_3= -{ G} (X,\phi){ \Box}\phi,
\nonumber\\
{\cal L}_4&=& G_4(X,\phi){ R}
+  G_{4, X}
\Big(
\big(\Box\phi\big)^2
-\phi_{\mu\nu} \phi^{\mu\nu}
\Big),
\nonumber\\
{{\cal L}}_5 &=& 
{G}_5(X,\phi){ G}_{\mu\nu}\phi^{\mu\nu}
\nonumber\\
&-&\frac{1}{6} {G}_{5, X}
\Big[
  ({\Box}\phi)^3
-3 (\Box\phi) \phi_{\mu\nu}\phi^{\mu\nu}
+2\phi_{\mu\alpha}\phi^{\alpha\nu}\phi^{\mu}_{\nu}
\Big],
\eea
where $P$, $G$,  $G_4$ and $G_5$ are free functions of both $X$ and $\phi$,
${\bar g}_{\mu\nu}$ is the matter frame metric
which is related to the gravity frame metric by the disformal relation \eqref{transfor},
$\Psi$ represents matter other than the scalar field,
and ${\cal L}_m$ represents the matter Lagrangian.
Following \cite{rule1},
through the disformal relation \eqref{transfor},
the gravitational action $S_g$ in \eqref{lag1}  can be rewritten in terms of the matter frame metric as
\bea
\label{lag2}
S_g=\int d^4 x\sqrt{-\bar g}\bar {\cal L}_g[\bar g,\phi] ,
\quad
\bar{\cal L}_g[\bar g,\phi]= \sum_{i=2}^5\bar{\cal L}_i[\bar g,\phi],
\eea
where $\bar{\cal L}_g$ is the Lagrangian of the gravity sector 
expressed with respect to the matter frame metric ${\bar g}_{\mu\nu}$
and $\bar{\cal L}_i$ is each Lagrangian of Horndeski's theory with respect to ${\bar g}_{\mu\nu}$. 
The relation between ${\cal L}_g$ and $\bar{\cal L}_g$ was obtained in \cite{rule1},
with the replacements $A\to \alpha$, $B\to \beta$ and $X\to-X$.
Readers who are interested in their explicit relation should refer to Appendix C of  Ref. \cite{rule1}.
The gravity and matter frames
are the natural extensions of the Einstein and Jordan frames 
in the case of the scalar-tensor theory with the nonminimal coupling 
of the scalar field to the Ricci scalar,
respectively,
where two frames are related by the conformal transformation.
Since the disformal transformation contains two free functions,
two more intermediate frames can also be defined as argued in \cite{dis_scr2,rule1},
which will not be considered in this paper.

In the gravity frame description \eqref{lag1},
by varying the action with respect to $g_{\mu\nu}$,
the gravitational equations of motion are obtained as
\bea
\label{einst}
0&=&E^{\mu\nu}+T_{(m)}^{\mu\nu},
\nonumber\\
E^{\mu\nu}&:=& 
\frac{2}{\sqrt{-g}}\frac{\delta \big(\sqrt{-g}{\cal L}_g\big)}{\delta g_{\mu\nu}},
\nonumber\\
T_{(m)}^{\mu\nu}&:=& \frac{2}{\sqrt{-g}}\frac{\delta \big(\sqrt{-\bar g(\phi)}{\cal L}_m\big)}{\delta g_{\mu\nu}},
\eea
where the gravitational tensor $E^{\mu\nu}$,
which reduces to the combination of Einstein tensor and the scalar field energy-momentum
in the case of the simplest Einstein-scalar theory,
is not divergent free and 
\bea
\label{cons}
-\nabla_\mu E^{\mu\nu}=\nabla_\mu T^{\mu\nu}_{(m)},
\eea
which represents the energy exchange between two sectors through the disformal coupling.

On the other hand,
in the matter frame description \eqref{lag2},
by varying the action with respect to ${\bar g}_{\mu\nu}$,
the gravitational equations of motion are obtained as 
\bea
\label{disf_eq}
0&=&{\bar E}^{\mu\nu}+{\bar T}_{(m)}^{\mu\nu},
\nonumber\\
{\bar E}^{\mu\nu}&:=& 
\frac{2}{\sqrt{-\bar g}}\frac{\delta \big(\sqrt{-\bar g}\bar {\cal L}_g\big)}{\delta \bar g_{\mu\nu}},
\nonumber\\
{\bar T}_{(m)}^{\mu\nu}&:=& \frac{2}{\sqrt{-\bar g}}\frac{\delta \big(\sqrt{-\bar g}{\cal L}_m\big)}{\delta \bar g_{\mu\nu}}.
\eea
In contrast to the previous gravity frame description \eqref{cons},
${\bar E}^{\mu\nu}$ and ${\bar T}^{\mu\nu}_{(m)}$ are separately conserved as
\bea
\label{rel}
-\bar \nabla_\mu {\bar E}^{\mu\nu}
={\bar \nabla}_\mu \bar T^{\mu\nu}_{(m)}=0,
\eea
where $\bar \nabla_{\mu}$ represents the covariant derivative with respect to 
the matter frame metric ${\bar g}_{\mu\nu}$.
The contravariant matter energy-momentum tensors defined in two frames are related by 
\bea
\label{dismat0}
T_{(m)}^{\mu\nu}
&=&\sqrt{\frac{\bar g}{g}} 
  \frac{\delta {\bar g}_{\alpha\beta}}{\delta g_{\mu\nu}} 
   {\bar T}_{(m)}^{\alpha\beta}
=
\alpha^3\sqrt{1-\frac{2X\beta}{\alpha}}
{\bar T}_{(m)}^{\mu\nu}.
\eea
Similarly, the mixed and covariant energy-momentum tensors are related by
\bea
\label{trans}
T_{(m)\nu}{}^\mu
&=&
\alpha^2\sqrt{1-\frac{2X\beta}{\alpha}}
\Big(\delta_\nu{}^\rho-\frac{\beta \phi_{\nu}\phi^{\rho}}{\alpha-2\beta X}\Big)
{\bar T}_{(m)\rho}{}^{\mu},
\nonumber\\
T_{(m)\mu\nu}
&=&
\alpha \sqrt{1-\frac{2X\beta}{\alpha}}
\Big(\delta_\mu{}^\rho-\frac{\beta \phi_{\mu}\phi^{\rho}}{\alpha-2\beta X}\Big)
\nonumber\\
&\times&\Big(\delta_\nu{}^\sigma-\frac{\beta \phi_{\nu}\phi^{\sigma}}{\alpha-2\beta X}\Big)
{\bar T}_{(m)\rho\sigma}.
\eea
Oppositely, 
\bea
\label{dismat}
\bar T_{(m)}^{\mu\nu}
&=&
\frac{1}{\alpha^3\sqrt{1-\frac{2X\beta}{\alpha}}}
{T}_{(m)}^{\mu\nu},
\nonumber\\
\bar {T}_{(m)\nu}{}^\mu
&=&
\frac{1}{\alpha^3\sqrt{1-\frac{2X\beta}{\alpha}}}
\big(\alpha\delta_\nu{}^\rho+\beta \phi_\nu\phi^\rho\big){T}_{(m)\rho}{}^{\mu},
\nonumber\\
\bar T_{(m)\mu\nu}
&=&
\frac{1}{\alpha^3\sqrt{1-\frac{2X\beta}{\alpha}}}
  \big(\alpha\delta_\mu{}^\rho+\beta \phi_\mu\phi^\rho\big)
\nonumber\\
&\times&
  \Big(\alpha \delta_\nu{}^\sigma+\beta\phi_\nu \phi^\sigma\Big)
{T}_{(m)\rho\sigma}.
\eea

So far, we have worked in the fixed coordinate system $x^{\mu}$.
In the homogeneous and isotropic cosmological background, however, 
the proper time coordinate in the gravity frame is not that in the matter frame (see the next section).
Thus, we introduce a new coordinate system $\hat x^{\mu}$ 
whose time component gives the proper time coordinate by
\bea
\label{dis_pro}
d\hat x^{\mu}:=\frac{\partial \hat x^{\mu}}{\partial x^\nu}dx^\nu.
\eea
The components of any tensor defined in the matter frame are related by
\bea
\label{dismat2}
\hat T^{\mu^1\nu^2\cdots \mu^m}{}_{\nu^1\nu^2\cdots \nu^n}
&=&\bar T^{\alpha^1\alpha^2\cdots \alpha^m}{}_{\beta^1\beta^2\cdots\beta^n}
\nonumber\\
&\times&
\Big(
\frac{\partial \hat x^{\mu^1}}{\partial x^{\alpha^1}}
\frac{\partial \hat x^{\mu^2}}{\partial x^{\alpha^2}}
\cdots
\frac{\partial \hat x^{\mu^m}}{\partial x^{\alpha^m}}
\Big)
\nonumber\\
&\times&
\Big(
\frac{\partial  x^{\beta^1}}{\partial \hat x^{\nu^1}}
\frac{\partial  x^{\beta^2}}{\partial \hat x^{\nu^2}}
\cdots
\frac{\partial  x^{\beta^n}}{\partial \hat x^{\nu^n}}
\Big).
\eea
In the hatted coordinate system,
the equations of motion in the matter frame \eqref{disf_eq} and \eqref{rel} are rewritten  as
\bea
0&=&{\hat E}^{\mu\nu}+{\hat T}_{(m)}^{\mu\nu},
\nonumber\\
{\hat E}^{\mu\nu}&:=& \frac{2}{\sqrt{-\hat g}}\frac{\delta \big(\sqrt{-\hat g}\hat {\cal L}_g\big)}{\delta \hat g_{\mu\nu}},
\nonumber\\
{\hat T}_{(m)}^{\mu\nu}&:=& \frac{2}{\sqrt{-\hat g}}\frac{\delta \big(\sqrt{-\hat g}{\cal L}_m\big)}{\delta \hat g_{\mu\nu}},
\eea
with
\bea
\label{rel3}
-\hat \nabla_\mu {\hat E}^{\mu\nu}={\hat \nabla}_\mu \hat T^{\mu\nu}_{(m)}=0,
\eea
where the covariant derivative $\hat \nabla_\mu$ is associated with $\hat g_{\mu\nu}$.

\section{Disformal transformation of gravitational perturbations}

In this section, we present the gauge-invariant gravitational perturbation variables in the 
gravity and matter frames and derive the relations between the corresponding perturbation variables in both frames.

\subsection{Disformal transformation of gravitational perturbations}

As the background solution in the gravity frame,
we consider the spatially-flat Friedmann-Lema\^itre-Robertson-Walker metric 
and the homogeneous scalar field
\bea
ds^2=g_{\mu\nu}dx^\mu dx^\nu=-dt^2+ a(t)^2\delta_{ij}dx^i dx^j,
\quad
\phi=\phi(t),
\eea
where $a(t)$ is the scale factor.
We then consider the scalar-type perturbations to the above background
\bea
\label{2}
ds^2&=&-(1+2A(t,x^i))dt^2+ 2a(t) \partial_i B(t,x^i) dt dx^i
\nonumber\\
&+& a(t)^2 \Big[\big(1-2\psi(t,x^i))\delta_{ij}+2\partial_i \partial_{j}E(t,x^i) \Big]dx^i dx^j,
\nonumber\\
\eea
and $\phi+\delta\phi$,
where
$A$, $B$, $\psi$ and $E$ are scalar-type metric perturbation variables.
Our notation of cosmological perturbations follows \cite{mfb}.
$\psi$ is often referred to as the curvature perturbation, 
as the intrinsic scalar curvature of the three-dimensional space is given by ${}^{(3)}R=\frac{4}{a^2}\Delta \psi$,
where $\Delta:= \delta^{ij}\partial_i\partial_j$.
In this paper, we will focus on the scalar-type perturbations,
as the vector- and tensor-type perturbations are not affected by the disformal transformation \eqref{transfor}.

Through the perturbation of the scalar field,
two functions in \eqref{transfor} are also perturbed as
\bea
\alpha\to \alpha+\alpha' \delta\phi,\quad
\beta\to \beta+\beta'\delta\phi,
\eea
and the metric in the matter frame is also perturbed.
Using \eqref{transfor}, then
the matter frame metric is given by 
\bea
\label{2}
d\bar s^2&=&
{\bar g}_{\mu\nu}dx^{\mu}dx^{\nu}
\nonumber\\
&=&-(\alpha-\beta \dot{\phi}^2+2\bar A(t,x^i))dt^2+ 2a(t) \partial_i \bar B(t,x^i) dt dx^i
\nonumber\\
&+& a(t)^2 \Big[\big(\alpha-2\bar \psi(t,x^i))\delta_{ij}+2\partial_i \partial_{j}\bar E(t,x^i) \Big]dx^i dx^j,
\nonumber\\
\eea
where the metric perturbations in the matter frame are given by  
\bea
\bar A&=&\alpha A+\frac{\alpha' \delta\phi} 
{2}-\frac{
\beta'\delta\phi}{2}\dot{\phi}^2-\beta\dot{\phi}\dot{\delta\phi},
\quad
\bar B=\alpha B+\frac{\beta\dot{\phi}}{a}\delta\phi,
\nonumber\\
\bar\psi&=& \alpha\psi -\frac{
\alpha'\delta\phi}{2},\quad
\bar E=\alpha E.
\eea
Then introducing the hatted coordinate system explained in the previous section by
\bea
\label{dpf}
d\hat t=\sqrt{\alpha-\beta \dot{\phi}^2}dt,\quad  \hat x^i= x^i,
\eea
the matter frame metric \eqref{2} is rewritten as 
\bea
d{\bar s}^2&=&
{\hat g}_{\mu\nu}d{\hat x}^{\mu}d{\hat x}^{\nu}
\nonumber\\
&=&-(1+2\hat A(\hat t,x^i))d\hat t^2+ 2\hat a(\hat  t) \partial_i\hat B(\hat t,x^i) d\hat t dx^i
\nonumber\\
&+& \hat a(\hat t)^2 \Big[\big(1-2\hat \psi(\hat t,x^i))\delta_{ij}+2\partial_i\partial_{j}\hat E(\hat t,x^i) \Big]dx^i dx^j,
\nonumber\\
\eea 
where $\hat a=\sqrt{\alpha}a$ is the scale factor in the matter frame, 
and 
the metric perturbations in the matter frames are given by
\bea
\label{dpf2}
\hat A&=&
\frac{\alpha A+\frac{\alpha'\delta\phi}{2}
-\frac{1}{2}\dot{\phi}^2
\beta'\delta\phi
 -\beta\dot{\phi}\dot{\delta\phi} }
{\alpha-\beta \dot{\phi}^2} ,
\quad
\hat B= \frac{B+\frac{\beta \dot{\phi}}{a\alpha}\delta\phi} {\sqrt{1-\frac{\beta}{\alpha} \dot{\phi}^2 }},
\nonumber\\
\hat \psi&=&\psi-\frac{\alpha'\delta\phi}{2\alpha},
\quad
\hat E=E.
\eea

Before closing this subsection, we comment on the (in)equivalence of the representative gauge conditions
under the disformal transformation. 
From \eqref{dpf2},
we find that the transformation of the spatial components of the metric perturbations, $\psi$ and $E$,
are not affected by the disformal component $\beta$,
while the remaining components, $A$ and $B$, are affected by $\beta$. 
Thus the synchronous gauges $A=B=0$ and $\hat  A=\hat B=0$ are not equivalent.
Similarly,
in contrast to the case of the pure conformal transformation $\beta=0$,
the longitudinal gauges $B=E=0$ and $\hat B=\hat E=0$ are also not equivalent under the disformal transformation.
On the other hand,
the scalar field perturbation $\delta\phi$ is invariant under the disformal transformation
and hence the comoving gauge to the scalar field $\delta\phi=0$ is unique.
As observables must be gauge-invariant,
in the next section the gauge-invariant gravitational perturbation variables are constructed as \cite{mfb,ks}.

\subsection{Gauge-invariant gravitational perturbations}

Under the gauge transformation $t\to t+\delta t$ and $x^i \to x^i+\delta^{ij} \partial_j \delta x$,
the metric and scalar field perturbation variables are transformed as
\bea
A&\to& A-\dot{\delta t},\quad
B\to B+ \frac{1}{a}\delta t-a\dot{\delta x},
\nonumber\\
\psi&\to& \psi +\frac{\dot{a}}{a}\delta t,\quad
E\to E-\delta x,\quad
\delta \phi\to \delta \phi-\dot{\phi}\delta t.
\eea
$B=E=0$ is the longitudinal gauge, 
$\psi=0$ is the spatially-flat gauge
and $\delta\phi=0$ is the comoving gauge to the scalar field, respectively.
Note that while in choosing the longitudinal gauge two gauge transformation functions $\delta t$ and $\delta x$ are completely fixed,
in choosing the spatially-flat and comoving gauges there is still the remaining gauge degree of freedom of $\delta x$.
Although in this paper we will call them `gauges',
they are also often called the spatially-flat and comoving `slices', respectively.

The representative gauge-invariant gravitational perturbation variables constructed in the gravity frame are given by 
\bea
\Phi&=&A-\frac{d}{dt}\Big[a^2\big(\dot{E}-\frac{B}{a}\big)\Big],
\nonumber\\
\Psi&=&\psi +a^2 \frac{\dot{a}}{a}\big(\dot{E}-\frac{B}{a}\big),
\nonumber\\
{\cal R}^{(\phi)}_c&=&\psi+\frac{1}{\dot{\phi}} \frac{\dot{a}}{a}\delta\phi.
\eea
Their counterparts in the matter frame are given by
\bea
\hat \Phi&=&\hat A-\frac{d}{d{\hat  t}}\Big[\hat a^2\big(\hat E_{,\hat t}-\frac{\hat B}{\hat a}\big)\Big],
\nonumber\\
\hat \Psi&=&\hat \psi +\hat a^2 \frac{\hat a_{,\hat t}}{\hat a}\big(\hat E_{,\hat t}-\frac{\hat B}{\hat a}\big),
\nonumber\\
\hat{\cal R}^{(\phi)}_c&=&\hat \psi+\frac{1}{\phi_{,\hat t}} \frac{\hat a_{,\hat t}}{\hat a}\delta\phi.
\eea
Using \eqref{dpf} and \eqref{dpf2},
we immediately find that the curvature perturbations in the comoving gauge to the scalar field are related by 
\bea
\label{invari}
\hat {\cal R}^{(\phi)}_c={\cal R}_c^{(\phi)}.
\eea 
Thus the comoving curvature perturbation to the scalar field is invariant under the disformal transformation
at least at the linear order of perturbations,
which is a generalization of the well-known conformal invariance of the same quantity \cite{cf1,cf3,cf4}.
In the case that only the scalar field is responsible for inflation and cosmological perturbations
and the comoving curvature perturbation is conserved on superhorizon scales \cite{nl,intr1,intr2}
\footnote{
As from \eqref{dpf} $\hat x^i=x^i$,
the fluctuation modes defined in both gravity and matter frames
can be labeled by the same comoving wavenumber $k$.
The notion of the `horizon crossing' of the mode with a given comoving wavenumber $k$ 
is, however, frame-dependent, as $\dot{a}\neq \hat{a}_{,\hat t}$. 
Nevertheless, the term `superhorizon'  can be commonly used for both frames,
for a sufficiently long wavelength mode of  $k\ll {\rm min} (\dot{a},{\hat a}_{,\hat t})$.
In the discussions below, by `superhorizon scales'
we mean the modes satisfying this condition.
},
the comoving curvature perturbation evaluated in any convenient disformally related frame 
during inflation gives the final observables.  
When there is matter disformally coupled to the scalar field, 
the curvature perturbation may not be conserved,
although the value of ${\cal R}_c^{(\phi)}$ is still frame-independent.

Similarly, the gauge-invariant metric perturbations in the longitudinal gauges are related by
\bea
\hat \Psi
&=&\Psi-\frac{1}{\alpha-\beta \dot{\phi}^2}
 \Big(\beta \dot{\phi}^2\frac{\dot{a}}{a}+\frac{\dot{\alpha}}{2}\Big)
\frac{\delta_g\phi}{\dot{\phi}},
\nonumber\\
\hat \Phi&=&
\frac{1}{\alpha-\beta \dot{\phi}^2}
\Big\{
  \alpha\Phi
\nonumber\\
&+&\frac{\dot{\alpha} (\alpha-2\beta \dot{\phi}^2)+\alpha\big(\dot{\phi}^2\dot{\beta}+2\beta \dot{\phi}\ddot{\phi}\big)}
         {2(\alpha-\beta \dot{\phi}^2)}
\frac{\delta_g\phi}{\dot{\phi}}
\Big\},
\nonumber\\
\eea
where we have defined the gauge-invariant perturbations of $Y$ in the longitudinal gauges by
\bea
\delta_g Y &=&\delta Y-a^2\dot{Y}\Big(\dot{E}-\frac{B}{a}\Big),
\nonumber\\
\hat \delta_g Y& =&\delta Y-\hat a^2Y_{,\hat t}\Big({\hat E}_{,\hat t}-\frac{\hat B}{\hat a}\Big).
\eea
The scalar field perturbations in the longitudinal gauges are related by 
\bea
\label{sca_g}
\hat \delta_g\phi=\delta\phi-\hat a^2\phi_{,\hat t}\Big(\hat{E}_{,\hat t}-\frac{\hat B}{\hat a}\Big)
=\frac{\alpha}{\alpha-\beta\dot{\phi}^2}\delta_g\phi.
\eea 
The scalar field perturbations in the spatially-flat gauges $\psi=0$ and $\hat \psi=0$ are defined by
\bea
\delta_\psi\phi&:=&\delta\phi+\frac{a}{\dot{a}}\dot{\phi}\psi
= \frac{\dot{a}}{a} \dot{\phi}  {\cal R}_c^{(\phi)},
\nonumber\\
\hat \delta_\psi\phi&=&
 \delta\phi+\frac{\hat a}{{\hat a}_{,\hat t}}\phi_{,\hat t}\hat \psi
= \frac{\hat a}{{\hat a}_{,\hat t}} \phi_{,\hat t} \hat {\cal R}_c^{(\phi)},
\eea
which are the primordial sources of cosmological perturbations on subhorizon scales in the single-field inflation models. 
Using \eqref{invari},
\bea
\label{sca_p}
\hat \delta_\psi \phi=\frac{\dot{a}}{a}\frac{{\hat a}}{{\hat a}_{,t}}\delta_\psi \phi.
\eea
Finally, the gauge-invariant combination
\bea
\Sigma^{(\phi)}:= A\dot{\phi}^2-\dot{\phi}\dot{\delta\phi}+\ddot{\phi}\delta\phi,
\eea
is related to the intrinsic entropy perturbation of the scalar field
$\Gamma^{(\phi)}:=\delta p^{(\phi)}-\frac{\dot{p}^{(\phi)}}{\dot{\rho}^{(\phi)}}\delta\rho^{(\phi)}$ \cite{intr1}.
This combination is disformally transformed as
\bea
\label{sca_i}
\hat\Sigma^{(\phi)}
:=
\hat A \phi_{,\hat t\hat t}-\phi_{,\hat t}\delta\phi_{,\hat t}+\phi_{,\hat t\hat t}\delta\phi 
=\frac{\alpha}{(\alpha-\beta\dot{\phi}^2)^2}
\Sigma^{(\phi)}.
\eea
As reviewed in Appendix,
in the single-field inflation models,
$\Sigma^{(\phi)}$ vanishes on superhorizon scales
if the comoving curvature perturbation is conserved on superhorizon scales, $\dot{\cal R}_c^{(\phi)}=0$ \cite{intr1,intr2}.
On the other hand,
if matter is disformally coupled,
the comoving curvature perturbation is not conserved on superhorizon scales
and hence $\Sigma^{(\phi)}$ is no longer suppressed.
Eqs. \eqref{sca_g}, \eqref{sca_p} and \eqref{sca_i} mean that 
the gauge-invariant scalar field perturbations are frame-independent, 
up to the 
rescalings by the background quantities.
In the next section, we will construct the gauge-invariant matter perturbation variables.

\section{Disformal transformation of matter perturbations}

In this section, we construct the gauge-invariant matter perturbation variables.

\subsection{Matter energy-momentum tensor}

In this section,
we assume that matter is composed of a set of noninteracting fluids
\bea
\label{king}
T_{(m)}^{\mu\nu}=\sum_a  T^{(a)\mu\nu},\quad
{\hat T}_{(m)}^{\mu\nu}=\sum_a {\hat T}^{(a)\mu\nu}.
\eea
According to \eqref{dismat} and \eqref{dismat2} 
the components of the energy-momentum tensors of the $(a)$-th component in both frames are
\bea
\label{dismat3}
\hat T^{(a)\mu\nu}
&=&
\frac{1}{\alpha^3\sqrt{1-\frac{2X\beta}{\alpha}}}
{T}^{(a)\alpha\beta}
 \frac{\partial \hat x^{\mu}}{\partial x^\alpha}
 \frac{\partial \hat x^{\nu}}{\partial x^\beta},
\nonumber\\
\nonumber\\
\hat {T}^{(a)}_{\nu}{}^\mu
&=&
\frac{1}{\alpha^3\sqrt{1-\frac{2X\beta}{\alpha}}}
\big(\alpha\delta_\rho{}^\alpha+\beta \phi_\rho\phi^\alpha\big)
{T}^{(a)}_{\alpha}{}^{\sigma}
\nonumber\\
&\times&
 \frac{\partial  x^{\rho}}{\partial \hat x^\nu}
 \frac{\partial \hat x^{\mu}}{\partial x^\sigma},
\nonumber\\
\hat T^{(a)}_{\mu\nu}
&=&
\frac{1}{\alpha^3\sqrt{1-\frac{2X\beta}{\alpha}}}
  \big(\alpha\delta_\rho{}^\alpha+\beta \phi_\rho\phi^\alpha\big)
\nonumber\\
&\times&
  \Big(\alpha \delta_\sigma{}^\beta+\beta\phi_\sigma \phi^\beta\Big)
T^{(a)}_{\alpha\beta}
 \frac{\partial  x^{\rho}}{\partial \hat x^\mu}
 \frac{\partial  x^{\sigma}}{\partial \hat x^\nu}.
\eea
As a further extension,
we may consider disformal coupling which depends on the component,
${\bar g}^{(a)}_{\mu\nu}=\alpha^{(a)} (\phi)g_{\mu\nu}+\beta^{(a)} (\phi)\phi_\mu\phi_\nu$.
In this paper, however, we are interested in how the gauge-invariant perturbations in
the gravity and matter frames are related, 
and do not consider such a generalized case.

We then derive the relation between the perturbed energy-momentum tensors defined in the gravity and matter frames.
We assume that in the gravity frame 
the energy-momentum tensor of the $(a)$-th component takes the following form
\bea
T^{(a)0}{}_0&=&-\rho^{(a)}-\delta\rho^{(a)}, \quad
T^{(a)i}{}_0=-\frac{\rho^{(a)}+p^{(a)}}{a}\partial^i v^{(a)},
\nonumber\\
T^{(a) i}{}_j
&=& \big(p^{(a)}+\delta p^{(a)}\big) \delta^i{}_j
+p^{(a)} 
\Big[\partial^i{}\partial_j
-\frac{1}{3}\delta^i{}_j\Delta
\Big]\Pi^{(a)},
\nonumber\\
&&
\eea
where $\rho^{(a)}$, $p^{(a)}$ are the background energy density and pressure of the $(a)$-th component,
and $\delta\rho^{(a)}$, $\delta p^{(a)}$, $v^{(a)}$ and $\Pi^{(a)}$
are the perturbed energy density, pressure, velocity and anisotropic stress
of the $(a)$-th component,
respectively.
The total matter energy-momentum tensor is given by 
\bea
T_{(m)}{}^{0}{}_{0}&=&-\rho-\delta\rho, \quad
T_{(m)}{}^{i}{}_{0}=-\frac{\rho+p}{a}\partial^i v,
\nonumber\\
T_{(m)}{}^{i}{}_{j}
&=& \big(p+\delta p\big) \delta^i{}_j
+p 
\Big[\partial^i{}\partial_j
-\frac{1}{3}\delta^i{}_j\Delta
\Big]
\Pi,
\eea
where from \eqref{king} 
\bea
\label{of0}
\rho=\sum_a \rho^{(a)},\quad
p=\sum_a p^{(a)},
\eea
and
\bea
\label{of1}
\delta \rho &=&\sum_{a}\delta \rho^{(a)},\quad
\delta p=\sum_{a}\delta p^{(a)},
\nonumber\\
(\rho+p) v
&=&\sum_a (\rho^{(a)}+p^{(a)}) v^{(a)},\quad
\nonumber\\
p \Pi&=&\sum_a p^{(a)}\Pi^{(a)}.
\eea
Similarly in the matter frame,
the energy-momentum tensor of the $(a)$-th component
is expressed  by
\bea
\hat T^{(a)0}{}_0&=&-\hat \rho^{(a)}-\delta \hat \rho^{(a)}, \quad
\hat T^{(a)i}{}_0=-\frac{\hat \rho^{(a)}+\hat p^{(a)}}{\hat a}\partial^i \hat v^{(a)},
\nonumber\\
\hat T^{(a)i}{}_j
&=& \big(\hat p^{(a)}+\delta \hat p^{(a)}\big) \delta^i{}_j
+\hat p^{(a)}
\Big[\partial^i{}\partial_j
-\frac{1}{3}\delta^i{}_j\Delta
\Big]
\hat \Pi^{(a)}.
\nonumber\\
\eea
The total matter energy-momentum tensor is also expressed as
\bea
\hat T_{(m)}{}^{0}{}_{0}&=&-\hat \rho-\delta \hat \rho, \quad
\hat T_{(m)}{}^{i}{}_{0}=-\frac{\hat \rho+\hat p}{\hat a}\partial^i \hat v,
\nonumber\\
\hat T_{(m)}{}^{i}{}_{j}
&=& \big(\hat p+\delta \hat p\big) \delta^i{}_j
+\hat p
\Big[\partial^i{}\partial_j
-\frac{1}{3}\delta^i{}_j\Delta
\Big]
\hat \Pi,
\eea
where from \eqref{king} 
\bea
\label{of2}
\hat \rho=\sum_a \hat \rho^{(a)},\quad
\hat p=\sum_a\hat  p^{(a)},
\eea
and
\bea
\label{of3}
\delta \hat \rho&=&\sum_{a}\delta \hat \rho^{(a)},\quad
\delta \hat p=\sum_{a}\delta \hat p^{(a)},
\nonumber\\
(\hat \rho+\hat p) \hat v
&=&\sum_a (\hat \rho^{(a)}+\hat p^{(a)}) \hat v^{(a)},\quad
\nonumber\\
\hat p\hat  \Pi&=&\sum_a \hat p^{(a)}\hat \Pi^{(a)}.
\eea
Using Eq. (\ref{dismat3}),
the background parts of the energy-momentum tensors of the $(a)$-th component
are transformed as  
\bea
\label{BG}
\hat \rho^{(a)}=f \rho^{(a)},
\quad
\hat p^{(a)} =\frac{\alpha}{\alpha-\beta \dot{\phi}^2} fp^{(a)},
\eea
where $f:=\frac{\sqrt{\alpha-\beta \dot{\phi}^2}}{\alpha^{\frac{5}{2}}}$.

Now we turn to the perturbation parts.
The perturbation parts of the energy-momentum tensors of the $(a)$-th component are related by 
\bea
\label{haha}
\frac{\delta\hat \rho^{(a)}}{\hat \rho^{(a)}}
&=&\frac{\delta\rho^{(a)}}{\rho^{(a)}}
-\frac{5\alpha'\delta\phi}{2\alpha}
\nonumber\\
&+&\frac{
\alpha'\delta\phi
-\dot{\phi}^2\beta'\delta\phi
 +2\beta (A\dot{\phi}^2-\dot{\phi}\dot{\delta\phi})}
         {2(\alpha-\beta\dot{\phi}^2)},
\nonumber \\
\frac{\delta\hat p^{(a)}}{\hat p^{(a)}}
&=&\frac{\delta p^{(a)}}{p^{(a)}}
-\frac{3\alpha'\delta\phi}{2\alpha}
\nonumber\\
&-&\frac{\alpha'\delta\phi
-\dot{\phi}^2\beta'\delta\phi +2\beta (A\dot{\phi}^2-\dot{\phi}\dot{\delta\phi})}
         {2(\alpha-\beta\dot{\phi}^2)},
\nonumber\\
\hat v^{(a)}&=&
\frac{\alpha\sqrt{1-\frac{\beta\dot{\phi}^2}{\alpha}}}
       {(\alpha-\beta\dot{\phi}^2)\rho^{(a)}+\alpha p^{(a)}}
\Big\{
(\rho^{(a)}+p^{(a)}) v^{(a)}
\nonumber\\
&-&\frac{\beta \dot{\phi}}{a}\frac{p^{(a)}}{\alpha-\beta\dot{\phi}^2}\delta\phi
-\frac{\beta\dot{\phi}^2 p^{(a)} }{\alpha-\beta\dot{\phi}^2}B
\Big\},
\nonumber\\
\hat \Pi^{(a)}&=&\Pi^{(a)}.
\eea
Thus the disformal transformation does not modify the structure of the perturbed energy-momentum tensor,
particularly a perfect fluid in the gravity frame ($\Pi^{(a)}=0$)
also corresponds to a perfect fluid in the matter frame ($\hat \Pi^{(a)}=0$). 
This is obvious from the fact that the background scalar field $\phi$ does not break the rotational invariance.

Clearly, from \eqref{haha},
the uniform density gauges $\delta\rho^{(a)}=0$ and $\delta\hat \rho^{(a)}=0$,
and 
the comoving gauges $v^{(a)}+B=0$ and $\hat v^{(a)}+\hat B=0$,
to the $(a)$-th component, are not equivalent between frames.

\subsection{Gauge-invariant matter perturbations}

Under the gauge transformation $t\to t+\delta t$ and $x^i \to x^i+\delta^{ij} \partial_j \delta x$,
the matter perturbation variables are transformed as
\bea
\delta\rho^{(a)}&\to& \delta\rho^{(a)}-\dot{\rho}^{(a)}\delta t,\quad
\delta p^{(a)}\to\delta p^{(a)}-\dot{p}^{(a)}\delta t,
\nonumber\\
 v^{(a)}&\to& v^{(a)}+a\dot{\delta x},\quad
\Pi^{(a)}\to  \Pi^{(a)}.
\eea
The gauge-invariant combinations of the matter energy-momentum tensor can also be 
constructed as those of the gravitational  perturbations \cite{mfb,ks}.
The curvature perturbations in the uniform energy density gauges
of the $(a)$-th component,
which have often appeared in the literature \cite{zeta1,malik},
are defined 
\bea
-\zeta^{(a)}:=\psi +\frac{\dot{a}}{ a}\frac{\delta \rho^{(a)}}{\dot{ \rho}^{(a)}},\quad
-\hat \zeta^{(a)}:= \hat \psi +\frac{{\hat a}_{,\hat t}}{\hat a}\frac{\delta\hat \rho^{(a)}}{{\hat \rho^{(a)}}_{,\hat t}},
\eea
Similarly, the curvature perturbations in the comoving gauges to the $(a)$-th component are given by
\bea
{\cal R}_c^{(a)}=\psi-\dot{a} (v^{(a)}+B),
\quad
{\hat {\cal R}}_c^{(a)}
=\hat\psi -{\hat a}_{,\hat t} (\hat v^{(a)}+\hat B).
\eea
The nonadiabatic pressure perturbations of the $(a)$-th component are given by 
\bea
\Gamma^{(a)}:= \delta p^{(a)}-\frac{{\dot p}^{(a)}}{\dot{\rho}^{(a)}}\delta\rho^{(a)},\quad
\hat \Gamma^{(a)}:=\delta \hat p^{(a)}-\frac{{\hat p^{(a)}}_{,\hat t}}{{\hat\rho^{(a)}}_{,\hat t}}\delta\hat \rho^{(a)}.
\eea
Comparing both frames, we find 
\bea
\label{qc}
-(\hat \zeta^{(a)}-\zeta^{(a)})
&=&
\frac{\frac{\dot{\rho}^{(a)}}{\rho^{(a)}}\frac{\dot{\alpha}}{2\alpha}-\frac{\dot a}{a}\frac{\dot{f}}{f}}
         { \frac{\dot{f}}{f}+\frac{\dot{\rho}^{(a)}}{\rho^{(a)}}}
\frac{1}{\dot{\rho}^{(a)}}\delta_\phi\rho^{(a)}
\nonumber\\
&+&\beta \frac{\frac{\dot a}{a}+\frac{\dot\alpha}{2\alpha}}
        { \frac{\dot{f}}{f}+\frac{\dot{\rho}^{(a)}}{\rho^{(a)}}}
\frac{\Sigma^{(\phi)}}
         {\alpha-\beta\dot{\phi}^2},
\nonumber \\
\hat{\cal R}^{(a)}_{c}-{\cal R}^{(a)}_{c}
&=&-\frac{\frac{\dot{\alpha}}{2}(\rho^{(a)}+p^{(a)})+\frac{\dot{a}}{a}\beta \rho^{(a)}\dot{\phi}^2}
         {(\alpha-\beta\dot{\phi}^2)\rho^{(a)}+\alpha p^{(a)}}
\frac{\delta_m^{(a)}\phi}{\dot{\phi}},
\nonumber\\
\frac{\hat\Gamma^{(a)}}{{\hat p}^{(a)}}
-\frac{\Gamma^{(a)}}{p^{(a)}}
&=&
-\beta \Big(1+\frac{{\hat p^{(a)}}_{,\hat t}}{{\hat \rho^{(a)}}_{,\hat t}}
            \frac{\hat \rho^{(a)}}{\hat p^{(a)}}
\Big)
\frac{\Sigma^{(\phi)}}
       {\alpha-\beta\dot{\phi}^2}
\nonumber\\
&-&
\Big(
\frac{\hat \rho^{(a)}}{\hat p^{(a)}}\frac{{\hat p^{(a)}}_{,\hat t}}{\hat\rho^{(a)}_{,\hat t}}
-\frac{\rho^{(a)}}{p^{(a)}}\frac{\dot{p}^{(a)}}{\dot{\rho}^{(a)}}
\Big)
\frac{\delta_\phi\rho^{(a)}}{\rho^{(a)}},
\eea
where we have defined
the perturbation of $Y$ in the comoving gauges to the scalar field 
\bea
\delta_\phi  Y&:=&\delta Y-\frac{\dot{Y}}{\dot{\phi}}\delta\phi,
\nonumber\\
\hat \delta_\phi  Y&:=&\delta Y-\frac{Y_{,\hat t}}{\phi_{,\hat t}}\delta\phi.
\eea 
and the perturbation of $Y$ in the comoving gauges to the $(a)$-th component,
\bea
\label{ma}
\delta^{(a)}_m Y&:=&\delta Y + a\dot{Y} (v^{(a)}+B),
\nonumber\\
{\hat \delta}^{(a)}_m Y&:=&\delta Y + \hat a Y_{,\hat t} ({\hat v}^{(a)}+ \hat B).
\eea
By definition, $\hat \delta_\phi  Y=\delta_\phi Y$.
In the limit of the purely conformal transformation, $\beta\to 0$,
the dependence on $\Sigma^{(\phi)}$ vanishes in \eqref{qc}.
The scalar field perturbations in the comoving gauges to the $(a)$-th component  are related by 
\bea
\hat \delta^{(a)}_m\phi
&=&\frac{\alpha (\rho^{(a)}+p^{(a)})}{(\alpha-\beta\dot{\phi}^2)\rho^{(a)}+\alpha p^{(a)}}\delta_m^{(a)}\phi.
\eea 
Thus the scalar field perturbations in the comoving gauges to the $(a)$-th component are also frame-independent,
up to the rescalings by the background dependent quantities,
as for the other gauge-invariant constructions of the scalar field perturbations.

The cosmological perturbations in the both frames are related as
\bea
&&\frac{\hat \delta_m\hat \rho^{(a)}}{\hat \rho^{(a)}}-\frac{\delta_m \rho^{(a)}}{\rho^{(a)}}
\nonumber\\
&=&\frac{\beta}{\alpha-\beta\dot{\phi}^2}
\Sigma^{(\phi)}
+\Big(
 \frac{\dot{f}}{f}
+\frac{\beta\dot{\phi}^2\rho^{(a)}}{(\alpha-\beta\dot{\phi}^2)\rho^{(a)}+\alpha p^{(a)}}
 \frac{\hat\rho^{(a)}_{,t}}{\hat \rho^{(a)}}
\Big)
\frac{\delta^{(a)}_m\phi}{\dot{\phi}},
\nonumber\\
&&\frac{\hat \delta_g \hat \rho^{(a)}}{\hat \rho^{(a)}}-\frac{\delta_g \rho^{(a)}}{\rho^{(a)}}
\nonumber\\
&=&\frac{\beta}{\alpha-\beta\dot{\phi}^2}
\Sigma^{(\phi)}
+\Big(
 \frac{\dot{f}}{f}
+\frac{\beta\dot{\phi}^2}{\alpha-\beta\dot{\phi}^2}
 \frac{{\hat \rho}^{(a)}_{,t}}{\hat \rho^{(a)}}
\Big)
\frac{\delta_g\phi}{\dot{\phi}}.
\eea
Thus their differences are determined by $\Sigma^{(\phi)}$ and the scalar field perturbations in the corresponding gauges.

The isocurvature perturbations between the $(a)$-th and $(b)$-th components are given by 
\bea
S^{(ab)}=3\big(\zeta^{(a)}-\zeta^{(b)}\big),
\quad
\hat S^{(ab)}=3\big(\hat \zeta^{(a)}-\hat \zeta^{(b)}\big),
\eea
and their difference is given by 
\bea
\hat S^{(ab)}-S^{(ab)}
&=&
-3\frac{\frac{\dot{\rho}^{(a)}}{\rho^{(a)}}\frac{\dot{\alpha}}{2\alpha}-\frac{\dot a}{a}\frac{\dot{f}}{f}}
         { \frac{\dot{f}}{f}+\frac{\dot{\rho}^{(a)}}{\rho^{(a)}}}
\frac{\delta_\phi\rho^{(a)} }
        {\dot{\rho}^{(a)}}
\nonumber\\
&+&3\frac{\frac{\dot{\rho}^{(b)}}{\rho^{(b)}}\frac{\dot{\alpha}}{2\alpha}-\frac{\dot a}{a}\frac{\dot{f}}{f}}
         { \frac{\dot{f}}{f}+\frac{\dot{\rho}^{(a)}}{\rho^{(a)}}}
\frac{\delta_\phi\rho^{(b)}}{\dot{\rho}^{(b)}}
\nonumber\\
&-&
3\frac{\beta  \big(\frac{\dot a}{a}+\frac{\dot\alpha}{2\alpha}\big)}{\alpha-\beta\dot{\phi}^2} 
  \frac{\frac{\dot{\rho}^{(b)}}{\rho^{(b)}}-\frac{\dot{\rho}^{(a)}}{\rho^{(a)}}}
       {(\frac{\dot{f}}{f}+\frac{\dot{\rho}^{(a)}}{\rho^{(a)}}) 
        (\frac{\dot{f}}{f}+\frac{\dot{\rho}^{(b)}}{\rho^{(b)}})} 
   \Sigma^{(\phi)}.
\eea
The curvature perturbations in the uniform energy density and 
comoving gauges to total matter are defined by
\bea
-\zeta &:=&\psi +\frac{\dot{a}}{a}\frac{\delta\rho}{\dot{\rho}},\quad
{\cal R}_c:= \psi -\dot{a}(\rho+p) (v+B),
\nonumber\\
-\hat \zeta &:=&\hat \psi +\frac{{\hat a}_{,\hat t}} {\hat a}\frac{\delta\hat \rho}{\hat \rho_{,\hat t}},\quad
\hat {\cal R}_c:= \hat \psi -{\hat a}_{,\hat t}(\hat \rho+\hat p) (\hat v+\hat B),
\eea
which from Eqs. \eqref{of0}, \eqref{of1}, \eqref{of2} and \eqref{of3}
are rewritten as
\bea
\label{tot_const}
\zeta&=&\sum_{a}\frac{\dot{\rho}^{(a)} } {\dot{\rho}}\zeta^{(a)},\quad
\hat \zeta=\sum_{a}\frac{\hat \rho_{,\hat t}^{(a)} } {\hat \rho_{,\hat t}}\hat \zeta^{(a)},
\nonumber\\
{\cal R}_c&=&\sum_{a}\frac{{\rho}^{(a)}+p^{(a)} } {\rho+p}{\cal R}_c^{(a)},\quad
\hat {\cal R}_c=\sum_{a}\frac{\hat {\rho}^{(a)}+\hat p^{(a)} } {\hat \rho+\hat p}
\hat {\cal R}_c^{(a)}.
\nonumber\\
\eea
Similarly, the scalar field perturbation in the comoving gauges to total matter are defined by
\bea
\delta_m\phi&:=& \delta\phi+ a\dot{\phi}(v+B),
\nonumber\\
\hat \delta_m \phi&:=& \delta\phi+ \hat a\phi_{,\hat t}(\hat v+\hat B),
\eea
which from \eqref{ma} are rewritten as
\bea
\delta_m\phi&=&\sum_{a}\frac{\rho^{(a)}+p^{(a)}}{\rho+p}\delta_m^{(a)}\phi,
\nonumber\\
\hat \delta_m\phi
&=&\sum_{a}\frac{\hat \rho^{(a)}+\hat p^{(a)}}{\hat \rho+\hat p}\hat \delta_m^{(a)}\phi.
\eea
The matter cosmological perturbations in the longitudinal gauges are given by 
\bea
\delta_g\rho=\sum_{a}\delta_g\rho^{(a)},\quad
\hat  \delta_g \hat \rho= \sum_{a}\hat \delta_g \hat \rho^{(a)}.
\eea
Finally, the total matter density perturbation in the comoving gauges to total matter is given by 
\bea
\delta_m\rho:=\delta\rho+ a(v+B)\dot{\rho},\quad
\hat \delta_m\hat \rho:=\delta\hat \rho+ \hat a(\hat v+\hat B){\hat \rho}_{,\hat t}.
\eea
Having the gauge-invariant perturbation variables and their differences between frames,
in the next section we will derive the gauge-invariant evolution equations of the curvature perturbations
in both frames.


\section{Evolution of curvature perturbations in the uniform energy density gauges}

In this section,
we derive the evolution equations of the curvature perturbation in the uniform energy density gauge of each component.
Our analysis in this section is based on the energy (non)conservation in the matter sector
and hence independent of the choice of the gravity sector.
In this section, we will work in the Fourier space and replace the spatial derivatives
with the comoving momentum as $\partial_i\to ik_i$ and $\Delta \to -k^2$ in the perturbation equations.

\subsection{Evolution of curvature perturbation in the matter frame}

In the matter frame $\hat g_{\mu\nu}$,
by definition matter is not directly coupled to the scalar field.
Under the assumption that 
there is no interaction between matter components,
the energy-momentum tensor of the $(a)$-th component of matter is conserved separately
$\hat \nabla_{\mu} {\hat T}^{(a)\mu}{}_{\nu}=0$.
Thus, the derivation of the evolution equation in the matter frame follows \cite{malik}.
The background part of the energy-momentum conservation law of the $(a)$-th component is given by
\bea
\label{queen}
\hat\rho^{(a)}_{,\hat t}
+3\frac{\hat a_{,\hat t}}{\hat a}
(\hat \rho^{(a)}+\hat p^{(a)})=0.
\eea
The perturbation part of the energy-momentum conservation law of the $(a)$-th component,
with use of the background relation \eqref{queen},
provides the evolution equation of the curvature perturbation in the uniform energy density gauges of the $(a)$-th component  
\bea
\label{cc}
\hat \zeta^{(a)}_{,\hat t}
=-\frac{1}{\hat \rho^{(a)}+\hat p^{(a)}} \frac{\hat a_{,\hat t}}{\hat a}\hat\Gamma^{(a)}
+\frac{1}{3} k^2 \Big( {\hat E}_{,\hat t}+\frac{\hat v^{(a)}}{\hat a}\Big),
\eea
where the $k^2$ terms in \eqref{cc} are suppressed on superhorizon scales.
If we analyze the cosmological dynamics in the matter frame,
after solving the evolution equation of the curvature perturbation for each component \eqref{cc}
with the gravitational and scalar field equations of motion,
the curvature perturbation in the uniform energy density gauge to total matter $\hat \zeta$
can be obtained from the second relation of \eqref{tot_const}.

\subsection{Evolution of curvature perturbation in the gravity frame}

The evolution of the curvature perturbation is more involved in the gravity frame $g_{\mu\nu}$
than in the matter frame ${\hat g}_{\mu\nu}$,
as the matter energy-momentum tensor is not conserved as \eqref{cons}.
Following \cite{dis_scr2}, 
we first derive the nonconservation of the energy-momentum tensor of matter,
\bea
\nabla_\mu E^{\mu}{}_\nu=-\nabla_\mu T_{(m)}^{\mu}{}_\nu
=Q \phi_\nu,
\eea 
where the coupling strength $Q$ in the right-hand side 
can be seen explicitly
from the divergence of the gravitational tensor
\bea
\nabla_\mu E^{\mu}{}_\nu
&=&
\Big\{
\frac{\partial {\cal L}_{g}
}{\partial\phi}
-\nabla_\mu\Big(\frac{\partial {\cal L}_g }{\partial \phi_\mu}\Big)
+\nabla_\alpha \nabla_\beta
 \Big(\frac{\partial {\cal L}_g }{\partial \phi_{\alpha\beta}}\Big)
\Big\}
\phi_\nu
\nonumber\\
&=&:Q\phi_\nu.
\eea 
As the terms in the curly bracket are equal to $\frac{\delta {\cal L}_g}{\delta \phi}$,
the equation of motion of the scalar field
$\frac{\delta {\cal L}_g}{\delta\phi}+\frac{\delta}{\delta\phi}\big(\sqrt{\frac{-\bar g}{-g}}{\cal L}_m\big)=0$
allows us to write the coupling term in terms of the variation of the matter Lagrangian as
\bea
\label{q}
Q=-\frac{1}{\sqrt{-g}}
\Big\{
\frac{\partial}{\partial\phi} (\sqrt{-\bar g}{\cal L}_m)
-\nabla_\mu \Big(\frac{\partial( \sqrt{-\bar g}{\cal L}_m )} {\partial \phi_{\mu}}\Big)
\Big\}.
\eea
Note that $\frac{\partial} {\partial \phi_{\alpha\beta}}(\sqrt{-\bar g}{\cal L}_m [\bar g,\Psi])=0$
and hence its contribution does not appear in \eqref{q},
as the matter frame metric ${\bar g}_{\mu\nu}$ defined in \eqref{transfor}
does not contain the second order derivative of the scalar field.
Then, using the chain rule, \eqref{transfor} and \eqref{einst}, 
each term in $Q$ can further be rewritten as
\bea
\label{chain}
\frac{\partial( \sqrt{-\bar g}{\cal L}_m )} {\partial \phi_{\mu}}
&=&\frac{\partial( \sqrt{-\bar g}{\cal L}_m )} {\partial g_{\rho\sigma}}
\frac{\partial g_{\rho\sigma}} {\partial {\bar g}_{\alpha\beta}}
\frac{\partial {\bar g}_{\alpha\beta}}{\partial \phi_\mu}
\nonumber\\
&=&\sqrt{-g}T_{(m)}^{\mu\nu}\frac{\beta}{\alpha}\phi_\nu,
\nonumber\\
\frac{\partial( \sqrt{-\bar g}{\cal L}_m )} {\partial \phi}
&=&\frac{\partial( \sqrt{-\bar g}{\cal L}_m )} {\partial g_{\rho\sigma}}
\frac{\partial g_{\rho\sigma}} {\partial {\bar g}_{\alpha\beta}}
\frac{\partial {\bar g}_{\alpha\beta}}{\partial \phi}
\nonumber\\
&=&\frac{1}{2}\sqrt{-g}T_{(m)}^{\mu\nu}\frac{1}{\alpha}
\big(\alpha' g_{\mu\nu}+\beta' \phi_\mu \phi_\nu\big).
\eea
Substituting \eqref{chain} into \eqref{q},
\bea
Q=\nabla_\rho\Big(\frac{\beta}{\alpha}T_{(m)}^{\rho\sigma}\phi_\sigma\Big)
-\frac{1}{2\alpha} T_{(m)}^{\rho\sigma} 
\big(\alpha' g_{\rho\sigma}+\beta' \phi_\rho\phi_\sigma\big).
\eea 
From \eqref{king}, the coupling term can be decomposed into the contributions of each component 
$Q=\sum_a Q^{(a)}$, where
\bea
Q^{(a)}:=  \nabla_\rho \Big(\frac{\beta}{\alpha} T^{(a)\rho\sigma}\phi_\sigma\Big)
-\frac{1}{2\alpha} T^{(a)\rho\sigma}\big(\alpha' g_{\rho\sigma}+\beta' \phi_\rho\phi_\sigma\big),
\eea
and the divergence of the energy-momentum tensor of the $(a)$-th component is given by 
\bea
\label{king3}
\nabla_\mu T^{(a)\mu}{}_\nu
&=&-Q^{(a)}\phi_\nu.
\eea
The background part of the energy-momentum nonconservation in the gravity frame \eqref{king3}
is given by
\bea
\label{bge}
&&\dot{\rho}^{(a)}+3\frac{\dot{a}}{a}(\rho^{(a)}+p^{(a)})
\nonumber\\
&=&\frac{\dot{\alpha}}{2\alpha}(\rho^{(a)}-3p^{(a)})
-\frac{\dot{\beta}}{2\alpha}\rho^{(a)}\dot{\phi}^2
\nonumber\\
&+&\frac{\beta}{\alpha}\dot{\rho}^{(a)}\dot{\phi}^2
+\frac{3\dot{a}}{a}\frac{\beta}{\alpha}\rho^{(a)}\dot{\phi}^2
+\rho^{(a)}\dot{\phi}\frac{d}{dt}\Big(\frac{\beta}{\alpha}\dot{\phi}\Big).
\eea
The perturbation part is given by
\begin{widetext}
\bea
\label{hiko}
&&\frac{d}{dt}\delta_\phi\rho^{(a)}
+3\frac{\dot{a}}{a}\big(\delta_\phi\rho^{(a)}+\delta_\phi p^{(a)}\big)
-3(\rho^{(a)}+p^{(a)})\dot{\cal R}_c^{(\phi)}
\nonumber\\
&=&\frac{\dot{\alpha}}{2\alpha}\big(\delta_\phi\rho^{(a)}-3\delta_\phi p^{(a)}\big)
-\frac{\dot{\beta}}{2\alpha}\rho^{(a)}\dot{\phi}^2
\Big(
-\frac{2\Sigma^{(\phi)} }{\dot{\phi^2}}
+\frac{\delta_\phi \rho^{(a)}}{\rho^{(a)}}
\Big)
\nonumber\\
&+&\dot{\phi}
\Big\{
3\frac{\dot{a}}{a}\frac{\beta}{\alpha}\rho^{(a)}\dot{\phi}
\Big(
-\frac{2\Sigma^{(\phi)}}{\dot{\phi^2}}
+\frac{\delta_\phi\rho^{(a)}}{\rho^{(a)}}
\Big)
+\frac{\beta}{\alpha}\rho^{(a)}\dot{\phi}
\frac{d}{dt}
\Big(
  \frac{\Sigma^{(\phi)} }{\dot{\phi^2}}
-3{\cal R}_c^{(\phi)}
\Big)
+\frac{d}{dt}
\Big[
\frac{\beta}{\alpha}\rho^{(a)}\dot{\phi}
\Big(
-\frac{2\Sigma^{(\phi)} }{\dot{\phi^2}}
+\frac{\delta_\phi \rho^{(a)}}{\rho^{(a)}}
\Big)
\Big]
\Big\}
\nonumber\\
&+&\frac{k^2}{a^2}
\Big[
\Big(1-\frac{\beta\dot{\phi}^2}{\alpha}\Big)
a(\rho^{(a)}+p^{(a)})\dot{\phi}\Big(v^{(a)}+a\dot{E}\Big)
-\frac{\beta\dot{\phi}^2}{\alpha}p^{(a)}
\frac{\delta_g\phi}{\dot{\phi}
}\Big],
\nonumber\\
&&\frac{d}{dt}
\Big[
(\rho^{(a)}+p^{(a)}) \frac{\delta_m^{(a)}\phi}{\dot{\phi}}
\Big]
+3\frac{\dot{a}}{a}
\Big[
(\rho^{(a)}+p^{(a)}) \frac{\delta_m^{(a)}\phi}{\dot{\phi}}
\Big]
+\delta_\phi p^{(a)}
+\big(\rho^{(a)}+p^{(a)}\big)\frac{\Sigma^{(\phi)}}{\dot{\phi}^2}
-\frac{2}{3}k^2p^{(a)}  \Pi^{(a)}
=0.
\eea
\end{widetext}
With the use of the relations between the gauge-invariant perturbations,
\bea
\delta_\phi\rho^{(a)}&=&-\frac{a \dot{\rho}^{(a)}}{\dot{a}}\big(\zeta^{(a)}+{\cal R}_c^{(\phi)}\big),\quad
\nonumber\\
\delta_\phi p^{(a)}
&=&\Gamma^{(a)}
-\frac{a \dot{p}^{(a)} }{\dot{a}}\big(\zeta^{(a)}+{\cal R}_c^{(\phi)}\big),
\nonumber\\
\delta_\phi \rho^{(a)}
&=&-\frac{\dot{\rho} ^{(a)}} {\dot{\phi}}
   \delta_{\rho^{(a)}}\phi,
\eea
where $\delta_{\rho^{(a)}}\phi:=\delta\phi-\frac{\dot \phi}{\dot\rho^{(a)}}\delta\rho^{(a)}$
is the scalar field perturbation in the uniform energy density gauge of the $(a)$-th component,
\eqref{bge} and \eqref{hiko},
the evolution equation of the curvature perturbation
 in the uniform energy density gauge of the $(a)$-th component  $\zeta^{(a)}$
is obtained as
\begin{widetext}
\bea
\label{ccc}
 \dot{\zeta}^{(a)}
&=&C^{(a)}_{1}\frac{\delta_{\rho^{(a)}}\phi} {\dot{\phi}}
+C^{(a)}_{2}
 \frac{d}{dt}\Big(\frac{\delta_{\rho^{(a)}}\phi} {\dot{\phi}}\Big)
+C_{3}^{(a)} \Sigma^{(\phi)}
+C_{4}^{(a)} \Gamma^{(a)}
\nonumber\\
&+&
\frac{\alpha-\beta\dot{\phi}^2}
        {3\big[\rho^{(a)}(\alpha-\beta\dot{\phi}^2)+p^{(a)}\alpha \big]}
\frac{k^2}{a^2}
\Big[
a(\rho^{(a)}+p^{(a)})\big(v^{(a)}+a\dot{E}\big)
-\frac{\beta\dot{\phi}^2  p^{(a)}} {\alpha-\beta\dot{\phi}^2}
\frac{\delta_g\phi}{\dot{\phi}}
\Big],
\eea
where
\bea
C^{(a)}_{1}
&=&
\frac{1}{6a^2\alpha \big(\rho^{(a)}(\alpha-\beta\dot{\phi}^2)+p^{(a)}\alpha\big)}
\Big\{
 2a^2\beta \dot{\alpha}\dot{\rho}^{(a)}\dot{\phi}^2
-6p^{(a)} \alpha^2 (\dot{a}^2-a\ddot{a})
\nonumber\\
&+&2\alpha^2
\Big(
-3\rho^{(a)}
 (\dot{a}^2-a\ddot{a})
+a 
\Big(
 3\dot{a}(\dot{\rho}^{(a)}+\dot{p}^{(a)})
+a\ddot{\rho}^{(a)}
\big)
\Big)
\nonumber\\
&+&\alpha
\Big[
3a^2\dot{p}^{(a)}\dot{\alpha}
-a^2\dot{\alpha}\dot{\rho}^{(a)}
+6\beta \rho^{(a)}\dot{a}^2\dot{\phi}^2
-6a\beta\dot{a}\dot{\rho}^{(a)}\dot{\phi}^2
-a^2\dot{\beta}\dot{\rho}^{(a)}\dot{\phi}^2
-6a\beta \rho^{(a)}\dot{\phi}^2\ddot{a}
-2a^2\beta \dot{\phi}^2\ddot{\rho}^{(a)}
-2a^2\beta \dot{\rho}^{(a)}\dot{\phi}\ddot{\phi}
\Big]
\Big\},
\nonumber\\
C^{(a)}_{2}&=&
\frac{1}{3a \big(\rho^{(a)}(\alpha-\beta\dot{\phi}^2)+p^{(a)}\alpha\big)}
\Big[
3p^{(a)} \alpha\dot{a}
+(3\rho^{(a)}\dot{a}+a\dot{\rho}^{(a)}) (\alpha-\beta\dot{\phi}^2)
\Big],
\nonumber\\
C_{3}^{(a)} 
&=&\frac{1}{3a\alpha \big(\rho^{(a)}(\alpha-\beta\dot{\phi}^2)+p^{(a)}\alpha\big)}
\Big[
 -a\alpha\rho^{(a)}\dot{\beta}
+2\beta
\big\{
a\rho^{(a)} \dot{\alpha}
-\alpha\big(3\rho^{(a)}\dot{a}+a\dot{\rho}^{(a)}\big)
\big\}
\Big],
\nonumber\\
C_4^{(a)}
&=&-\frac{2\alpha\dot{a}+a\dot{\alpha}}
{2a\big(\rho^{(a)}(\alpha-\beta\dot{\phi}^2)+p^{(a)}\alpha\big)}.
\eea
\end{widetext}
The terms in the second line in \eqref{ccc} which are proportional to $k^2$
are suppressed on superhorizon scales.
As a quick check, 
in the limit of $\alpha\to1$ and $\beta\to 0$ where two frames coincide,
with use of the energy conservation law,
\eqref{ccc} reduces to \eqref{cc} without `hat'.
Because of the disformal coupling of matter to the scalar field,
in addition to the nonadiabatic pressure $\Gamma^{(a)}$,
$\delta_{\rho^{(a)}} \phi$ and $\Sigma^{(\phi)}$ are also sourcing the curvature perturbation on superhorizon scales. 
If we analyze the cosmological dynamics in the gravity frame,
solving \eqref{ccc} for each component
with the gravitational and scalar field equations of motion,
the curvature perturbation in the uniform matter energy density gauge in the gravity frame $\zeta$ 
is obtained from the first relation of \eqref{tot_const}.

Let  us check the adiabaticity and 
the condition of the conservation of the curvature perturbations on superhorizon scales
in both frames.
Using \eqref{qc},
we find that when the nonadiabatic pressure of the $(a)$-th component 
in the matter frame vanishes, $\hat \Gamma^{(a)}=0$,
and hence the curvature perturbation in the uniform energy density gauge of the $(a)$-th component 
in the matter frame is conserved on superhorizon scales, ${\hat \zeta}_{,\hat t}^{(a)}=0$,
the nonadiabatic pressure in the gravity frame is given by 
\bea
\Gamma^{(a)}
&=&
p^{(a)}
\beta \Big(1+\frac{{\hat p^{(a)}}_{,\hat t}}{{\hat \rho^{(a)}}_{,\hat t}}
            \frac{\hat \rho^{(a)}}{\hat p^{(a)}}
\Big)
\frac{\Sigma^{(\phi)}}
       {\alpha-\beta\dot{\phi}^2}
\nonumber\\
&-&\frac{p^{(a)}}{\rho^{(a)}}
  \dot{\rho}^{(a)}
\Big(
\frac{\hat \rho^{(a)}}{\hat p^{(a)}}\frac{{\hat p^{(a)}}_{,\hat t}}{\hat\rho^{(a)}_{,\hat t}}
-\frac{\rho^{(a)}}{p^{(a)}}\frac{\dot{p}^{(a)}}{\dot{\rho}^{(a)}}
\Big)
\frac{\delta_{\rho^{(a)}}\phi} {\dot{\phi}},
\eea
which does not vanish in general,
meaning that the adiabaticity condition for the $(a)$-th component
does not hold in the gravity frame.
On superhorizon scales,
\eqref{ccc} reduces to
\bea
\label{super}
\dot{\zeta}^{(a)}
&\approx& 
\Big[C^{(a)}_1
-\frac{p^{(a)}}{\rho^{(a)}}
  \dot{\rho}^{(a)}
\Big(
\frac{\hat \rho^{(a)}}{\hat p^{(a)}}\frac{{\hat p^{(a)}}_{,\hat t}}{\hat\rho^{(a)}_{,\hat t}}
-\frac{\rho^{(a)}}{p^{(a)}}\frac{\dot{p}^{(a)}}{\dot{\rho}^{(a)}}
\Big)
C_4^{(a)}
\Big]
\frac{\delta_{\rho^{(a)}}\phi }{\dot{\phi}}
\nonumber\\
&+&C^{(a)}_{2}
 \frac{d}{dt}\Big(\frac{\delta_{\rho^{(a)}}\phi} {\dot{\phi}}\Big)
\nonumber\\
&+&\Big[
C_3^{(a)}
+\frac{\beta p^{(a)}}{\alpha-\beta\dot{\phi}^2}
\Big(1+
\frac{{\hat p}^{(a)}_{,t} } {{\hat \rho}^{(a)}_{,t}}
\frac{\hat \rho^{(a)} } {{\hat p}^{(a)}}
\Big)
C_4^{(a)}
\Big]
 \Sigma^{(\phi)},
\eea
meaning that the curvature perturbation in the uniform energy density gauge 
in the gravity frame is not conserved,
even if the corresponding curvature perturbation is conserved in the matter frame.

The arguments so far hold for any choice of the gravity sector,
as the evolution equations of the curvature perturbations in the uniform energy density gauges \eqref{cc}
and \eqref{ccc} are derived solely from the energy (non)conservation in the matter sector \cite{malik}.
Of course, in order to obtain the closed sets of the equations of motion in each frame,
the gravitational and scalar field equations of motion should be derived
for a given gravity sector.
In evaluating the observables, we need to choose a particular frame where the equations of motion are solved.
As it is more convenient to analyze the cosmological dynamics in the Einstein frame
than in the Jordan frame in the case of the scalar-tensor theory with the nonminimal coupling
of the scalar field to the Ricci scalar,
it would also be more convenient to analyze the cosmological dynamics in the gravity frame than in the matter frame.
In the gravity frame,
the evolution equations of the curvature perturbations \eqref{ccc} are then solved 
by combining the gravitational and scalar field equations of motion. 
As in the theories considered in this paper
CMB photons follow null geodesics associated with the matter frame,
the curvature perturbation in the uniform matter energy density gauge
in the matter frame $\hat \zeta$ should be finally obtained. 
Once the curvature perturbation in the gravity frame $\zeta^{(a)}$ is evaluated,
the curvature perturbation in the matter frame $\hat \zeta^{(a)}$
is evaluated 
via the frame-transformation rule \eqref{qc} or equivalently
\bea
-{\hat \rho}^{(a)}_{,t} {\hat \zeta}^{(a)}
&=&-f\dot{\rho}^{(a)}\zeta^{(a)}
+\dot{f} \rho^{(a)}{\cal R}_c^{(\phi)}
+f\frac{\dot{\alpha}}{2\alpha}\delta_\phi\rho^{(a)}
\nonumber\\
&+&\beta f \rho^{(a)}
\Big(\frac{\dot{a}}{a}+\frac{\dot{\alpha}}{2\alpha}\Big)
\frac{\Sigma^{(\phi)}}{\alpha-\beta\dot{\phi}^2}.
\eea
The curvature perturbation on the uniform matter energy density gauge in the matter frame $\hat \zeta$
is then evaluated by the second relation of \eqref{tot_const}.


\section{Conclusions}

As the extension of the case of the well-studied conformal transformation,
in this paper we have derived the disformal transformation rules of the gauge-invariant cosmological perturbation variables.
The disformal transformation has been argued in terms of applications to the specific cosmological problems,
and more recently has been focused as the way to transform a class of Horndeski's theory to another.
We have considered a class of 
the scalar-tensor theory
where the gravity and matter frame metrics
are related by the disformal transformation \eqref{transfor}.
We have assumed a class of Horndeski's theory as the gravity sector
and noninteracting fluids disformally coupled to the scalar field as the matter sector.
We have also assumed that the disformal coupling to matter is common,
in the sense that the form of the coupling is independent of the matter components.

Concerning the perturbations associated with the gravity sector, it is straightforward to confirm that
the curvature perturbation in the comoving gauge to the scalar field is invariant
under the disformal transformation as well as the conformal transformation.
In case that only the scalar field is responsible both for inflation and cosmological perturbations,
the frame-independent curvature perturbation comoving to the scalar field is conserved on superhorizon scales
and gives the final predictions from inflation.
Having the disformally invariant curvature perturbation, it can be evaluated in the most convenient frame,
as usually done in the Einstein frame in the case of the scalar-tensor theory with the nonminimal coupling 
of the scalar field to the Ricci scalar.
Since the disformal transformation \eqref{transfor} is induced by the scalar field,
the vector- and tensor-type perturbations are invariant under the disformal transformation.
Combining with the fact that the comoving curvature perturbation is invariant,
the tensor-to-scalar ratio is also invariant in the single-field inflation models.

Concerning the perturbations associated with the matter sector,
in the matter frame the curvature perturbation in the uniform matter energy density gauge 
is conserved on superhorizon scales,
if the nonadiabatic pressure defined in the matter frame vanishes. 
On the other hand, in the gravity frame the curvature perturbation in the uniform matter energy density gauge 
is not conserved,  even if the nonadiabatic pressure defined in the gravity frame vanishes.
As the evolution equations of the curvature perturbation
were obtained from the energy (non)conservation in the matter sector,
the evolutions equations of the curvature perturbation hold
for any choice of the gravity sector.

Before closing this paper, we would like to mention the related works.
First, we emphasize that the results presented in this paper 
and the disformal invariance of vector and tensor perturbations
would apply only to the linear perturbations.
Therefore, the analysis beyond the linear order would be very important.
There would also be various extensions of the present work:
more general disformal transformation \eqref{transfor0}
which would allow for the healthy scalar-tensor theory beyond Horndeski,
component-dependent disformal couplings,
disformal transformations by other field species \cite{rule2}
and so on.
We hope to come back to these issues in our future publications.

\appendix

\section*{Acknowledgement}
This work was supported by the FCT Portugal through Grant No. SFRH/BPD/88299/2012.

\section{Intrinsic entropy perturbation in the single field inflation models}

In this appendix,
we will confirm that the gauge-invariant perturbation $\Sigma^{(\phi)}$
is suppressed on superhorizon scales in the general single-field models of inflation
in which the gravity action is given by the Einstein gravity,
i.e., $G_4=\frac{1}{2\kappa^2}$ and $G_5=0$ in \eqref{gal_acc}:
\bea
S=\int  d^4 x\sqrt{-g}\Big[\frac{1}{2\kappa^2}R+P(X,\phi)-G(X,\phi)\Box\phi\Big].
\eea
This theory involves the single-field slow-roll and k-inflation models for $G=0$ \cite{inf1}
and models with the generalized cubic galileon for $G\neq 0$ \label{hor_inf1}. 
For $G=0$, the suppressed intrinsic entropy perturbation was argued in \cite{intr1}.

Varying the action with respect to the metric $g_{\mu\nu}$, 
the gravitational equation of motion is obtained as
\bea
\label{eins}
G_{\mu\nu}&=&\kappa^2 T_{(\phi)\mu\nu},
\nonumber\\
T^{(\phi)}_{\mu\nu}&=& Pg_{\mu\nu} +P_X\phi_\mu\phi_\nu
-G_X \Box \phi \phi_\mu\phi_\nu
+ g^{\rho\sigma} J_\rho \phi_\sigma g_{\mu\nu}
\nonumber\\
&-&\Big(
J_{\mu}\phi_\nu + J_{\nu}\phi_\mu
\Big),
\nonumber\\
J_\rho&:=& G_\phi \phi_\rho-G_X g^{\alpha\beta}\phi_\alpha\phi_{\beta\rho}.
\eea
The background equation of motion of the scalar field is given by
\bea
0&=&\ddot{\phi}
+\Big(\frac{3\dot{a}}{a}+\frac{P_{X\phi}\dot{\phi} + P_{XX}\dot{\phi}\ddot{\phi}}{P_X}\Big)\dot{\phi}
-\frac{P_\phi}{P_X}
\nonumber\\
&+&2\Big(-G_\phi+3\frac{\dot{a}}{a}G_X\dot{\phi}\Big)\ddot\phi
\nonumber\\
&+&3\Big(G_{X\phi}+G_{XX}\ddot{\phi} \Big)\frac{\dot{a}}{a}\dot{\phi}^3
+3G_X \Big(\frac{d}{dt}\Big(\frac{\dot{\phi}}{\phi}\Big)+3\frac{\dot{a}^2}{a^2}\Big)\dot{\phi}^2
\nonumber\\
&-&
\Big(
  G_{\phi\phi}\dot{\phi}
+G_{X\phi} \dot{\phi}\ddot{\phi}
+6\frac{\dot{a}}{a} G_\phi
\Big)
\dot{\phi}.
\eea
The background part of the Laplacian is given by
$\Box\phi=-\ddot{\phi}-3\frac{\dot{a}}{a}\dot{\phi}$.
The perturbation part of the Hamiltonian and momentum constraints obtained from
 \eqref{eins} is given by
\bea
\label{pert_eins}
\frac{3\dot{a}}{a}\big(\dot{\psi}+\frac{\dot{a}}{a}A\big)
+\frac{k^2}{a^2}
\Big[\psi +a\dot{a} \Big(\dot{E}-\frac{B}{a}\Big)\Big]
&=&-\frac{\kappa^2}{2}\delta\rho^{(\phi)},
\nonumber\\
\dot{\psi}+\frac{\dot{a}}{a}A &=&-\frac{\kappa^2}{2}\delta q^{(\phi)},
\eea
where
\begin{widetext}
\bea
\delta\rho^{(\phi)}
&=&-\big(P_\phi \delta\phi + P_X \delta X\big)
+2 XP_X \Big(
  \frac{P_{X\phi}}{P_X}\delta\phi
+ \frac{P_{XX}}{P_X}\delta X
+2\frac{\dot{\delta\phi}}{\dot{\phi}}
-2A
\Big)
\nonumber\\
&-&G_X \big(\Box \phi \big)\dot{\phi}^2
\Big(
  \frac{G_{X\phi}}{G_X}\delta\phi
+  \frac{G_{XX}}{G_X }\delta X
+\frac{\delta (\Box\phi)}{\Box\phi}
-2A
+2\frac{\dot{\delta\phi}}{\dot{\phi}}
\Big)
\nonumber\\
&-&\dot{\phi}^2 \big(G_\phi+ G_X\ddot{\phi}\big)
\Big[
-2A+\frac{\dot{\delta\phi}}{\dot{\phi}}
+\frac{G_\phi}{G_\phi+ G_X\ddot{\phi}}
\Big( 
  \frac{G_{\phi\phi}}{G_{\phi}}\delta\phi
+\frac{G_{X\phi}}{G_{\phi}}\delta X
+\frac{\dot{\delta\phi}}{\dot{\phi}}
\Big)
\nonumber\\
&+&\frac{G_X\ddot{\phi}}{G_\phi+ G_X\ddot{\phi}}
\Big( 
  \frac{G_{X\phi}}{G_X}\delta\phi
+\frac{G_{XX}}{G_X}\delta X
-2A 
+\frac{\dot{\delta\phi}}{\dot{\phi}}
+\frac{\ddot{\delta\phi}} {\ddot{\phi}}
-\frac{\dot{A}\dot{\phi}}{\ddot{\phi}}
\Big)
\Big],
\nonumber\\
\delta q^{(\phi)}&=&
-2 X P_X \frac{\delta\phi}{\dot{\phi}}
+
\Big[
G_X(\Box\phi)\dot{\phi}\delta\phi
+\dot{\phi}\Big(G_\phi\delta\phi +G_X \dot{\phi}\big(\dot{\delta\phi}-A\dot{\phi}\big)\Big)
+\big(G_\phi +G_X\ddot{\phi} \big)\dot{\phi}\delta\phi
\Big],
\eea
\end{widetext}
with
$\delta X=-A\dot{\phi}^2 +\dot{\phi}\dot{\delta\phi}$,
and 
$\delta (\Box\phi)
=-\Big(\ddot{\delta\phi}+3\frac{\dot{a}}{a}\dot{\delta\phi}\Big)
+2A \Big(\ddot{\phi}+3\frac{\dot{a}}{a}\dot{\phi}\Big)
+\dot{A}
+3\dot{\psi}
+\frac{1}{a^2}
\Delta \delta_g\phi$.
Combining these constraints, we obtain
$-\frac{k^2}{a^2}\Psi=\frac{1}{2}\tilde \delta_m \rho^{(\phi)}$
where
\bea
\label{tilder}
\tilde \delta_m \rho^{(\phi)}
&=&
\Big\{
-\big(P_X+ 2XP_{XX}\big)
-\frac{\dot{\phi}^2}{\ddot{\phi}}G_{\phi\phi}
+3\frac{\dot{a}}{a}\frac{\dot{\phi}^3}{\ddot{\phi}} G_{X\phi}
\nonumber\\
&-&6\frac{\dot{a}}{a}\frac{\dot{\phi}}{\ddot{\phi}} G_\phi
+3\frac{\dot{\phi}}{a^2\ddot{\phi}}
 \big(2\dot{a}^2\dot{\phi}+a\dot{\phi}\ddot{a}-a\dot{a}\ddot{\phi}\big)G_X
\Big\}
\Sigma^{(\phi)}
\nonumber\\
&-&3\dot{\phi}^3 G_X\dot{{\cal R}}_c^{(\phi)}.
\eea
The third line of the right-hand side of \eqref{tilder} vanishes 
when the comoving curvature perturbation is conserved $\dot{\cal R}^{(\phi)}_c=0$ \cite{intr2}.
Thus in this case $\Sigma^{(\phi)}$ is also suppressed on superhorizon scales.

It is more involved to check whether $\Sigma^{(\phi)}$
is suppressed for the more general gravitational theory where the gravity action also contains
nonminimal couplings to the curvature tensors with $G_4\neq {\rm const}$ and $G_{5}\neq 0$.
But, with the (inverse) disformal transformation
$g_{\mu\nu}=\frac{1}{\alpha} \big({\bar g}_{\mu\nu}-\beta\phi_{|\mu}\phi_{|\nu}\big)$,
it is possible to rewrite the Einstein gravity with the form of the disformal coupling  
\begin{widetext}
\bea
&&\int d^4x \sqrt{-g}\Big[R+\cdots\Big]
=
 \int d^4 x\sqrt{-\hat g} 
\Big\{
{\hat G}_4(\hat X,\phi) {\hat R}
+{\hat G}_{4,\hat X} (\hat X,\phi)
\Big( (\hat \Box\phi)^2-
\big(\hat \nabla_\mu\hat \nabla_\nu \phi) (\hat \nabla^\mu\hat \nabla^\nu \phi\big)\Big)
+\cdots 
\Big\}.
\eea
\end{widetext}
with ${\hat G}_4(\hat X,\phi) :=\frac{1}{\alpha}    (1-2\beta\hat X)^{\frac{1}{2}}$.
This involves the model with the nonminimal coupling
${\hat G}_4=\frac{1}{\alpha(\phi)}$ for $\beta=0$ \cite{cf1,cf3,cf4,nm_inf}
and DBI galileon for $\beta \neq 0$ \cite{dis_scr2}.
As $\Sigma^{(\phi)}$ in the original frame is suppressed by $k^2$, 
through the relation \eqref{sca_i}
$\hat \Sigma^{(\phi)}$ is also suppressed on superhorizon scales.


\end{document}